\documentclass[aps,prd,amsmath,amssymb,preprint,superscriptaddress]{revtex4}

\usepackage{array} 
\usepackage{amssymb}
\usepackage{graphics,graphpap}
\usepackage{graphicx}
\usepackage{color}
\usepackage{graphicx}
\usepackage{dcolumn}
\usepackage{epsfig}
\usepackage{epstopdf}
\usepackage{bm}
\usepackage{amsmath}
\usepackage{amsfonts}
\usepackage{textcomp}
\usepackage{setspace}

\newcommand{\beq}{\begin{eqnarray}}
\newcommand{\eeq}{\end{eqnarray}}

\newcommand{\bmp}{\noindent\begin{minipage}{16cm}}
\newcommand{\emp}{\end{minipage}\vskip 7mm} 

\def\drawbox#1#2{\hrule height#2pt
        \hbox{\vrule width#2pt height#1pt \kern#1pt
              \vrule width#2pt}
              \hrule height#2pt}

\def\Asym#1#2{\vcenter{\vbox{\drawbox{#1}{#2}
              \kern-#2pt 
              \drawbox{#1}{#2}}}}

\def\simge{\mathrel{%
   \rlap{\raise 0.511ex \hbox{$>$}}{\lower 0.511ex \hbox{$\sim$}}}}

\def\simle{\mathrel{
   \rlap{\raise 0.511ex \hbox{$<$}}{\lower 0.511ex \hbox{$\sim$}}}}

\def\s#1{\setbox0=\hbox{$#1$}%
\rlap{\ifdim\wd0>.7em\kern.22\wd0\else\kern.1\wd0\fi /}#1}

\newcommand{\ie}{i.e.}
\newcommand{\eg}{e.g.}

\newcommand{\ud}{{\rm d}}

\newcommand{\todo}[1]{(\textbf{TODO:} #1)}
\renewcommand{\todo}[1]{}

\def\slc#1{\setbox0=\hbox{$#1$}           
    \dimen0=\wd0                                 
    \setbox1=\hbox{/} \dimen1=\wd1               
    \ifdim\dimen0>\dimen1                        
       \rlap{\hbox to \dimen0{\hfil/\hfil}}      
       #1                                        
    \else                                        
       \rlap{\hbox to \dimen1{\hfil$#1$\hfil}}   
       /                                         
    \fi}

\begin{document}

\title{Higgs Dark Matter in UEDs:\\A Good WIMP with Bad Detection Prospects}

\date{\today}

\author{Henrik Melb\'eus}
\email{melbeus@kth.se}

\affiliation{Department of Theoretical Physics, School of Engineering Sciences, KTH Royal Institute of Technology -- AlbaNova University Center, Roslagstullsbacken 21, 106 91 Stockholm, Sweden}

\author{Alexander Merle}
\email{A.Merle@soton.ac.uk}

\affiliation{Department of Theoretical Physics, School of Engineering Sciences, KTH Royal Institute of Technology -- AlbaNova University Center, Roslagstullsbacken 21, 106 91 Stockholm, Sweden}
\affiliation{Physics and Astronomy, University of Southampton, Highfield, Southampton SO17 1BJ, United Kingdom}

\author{Tommy Ohlsson}
\email{tommy@theophys.kth.se}

\affiliation{Department of Theoretical Physics, School of Engineering Sciences, KTH Royal Institute of Technology -- AlbaNova University Center, Roslagstullsbacken 21, 106 91 Stockholm, Sweden}

\begin{abstract}

We study the first Kaluza--Klein excitation of the Higgs boson in universal extra dimensions as a dark matter candidate. The first-level Higgs boson could be the lightest Kaluza--Klein particle, which is stable due to the conservation of Kaluza--Klein parity, in non-minimal models where boundary localized terms modify the mass spectrum. We calculate the relic abundance and find that it agrees with the observed dark matter density if the mass of the first-level Higgs boson is slightly above 2~TeV, not considering coannihilations and assuming no relative mass splitting among the first-level Kaluza--Klein modes. In the case of coannihilations and a non-zero mass splitting, the mass of the first-level Higgs boson can range from 1~TeV to 4~TeV. We study also the prospects for detection of this dark matter candidate in direct as well as indirect detection experiments. Although the first-level Higgs boson is a typical weakly interacting massive particle, an observation in any of the conventional experiments is very challenging.

\end{abstract}

\maketitle

\section{\label{sec:intro}Introduction}

The particle identity of dark matter (DM) is one of the most important questions in particle physics, both from a theoretical as well as an experimental point of view. Since none of the particles in the Standard Model (SM) of particle physics can make up the DM, the problem points to new physics beyond the SM. The most popular class of DM candidates is weakly interacting massive particles (WIMPs), which are particles that have weak interactions and masses in the GeV to TeV range.

One of the possibilities for new physics beyond the SM is provided by models with extra spatial dimensions. In particular, in models with universal extra dimensions (UED)~\cite{Appelquist:2000nn}, the conservation of Kaluza--Klein (KK) parity ensures the stability of the lightest KK particle (LKP), which may be a viable WIMP DM candidate \cite{Servant:2002aq}. In the five-dimensional UED model, the internal space has to be compactified on the orbifold $S^1 / {\mathbb Z}_2$ in order to give chiral fermions at the level of the zero modes and to avoid the existence of massless fifth components of the gauge fields. At the excited KK levels, each fermion field $f$ in the SM is replaced by two Dirac fermions, $f_{\rm D}$ and $f_{\rm S}$, having the quantum numbers of the corresponding left- and right-handed SM fermions, respectively. In addition, each of the KK excitations of the electroweak gauge bosons obtains a mass by eating a scalar that is a combination of the excitations of the SM Higgs field and the fifth components of the gauge bosons. In the limit of a large compactification scale $R^{-1}$ in comparison to the Higgs vacuum expectation value, these Goldstone bosons are mainly composed of the fifth components of gauge bosons, and hence, the KK excitations of the charged component $H^\pm$ and of the pseudo-scalar $A^0$ are physical particles.

The mass spectrum in the UED model, and thus the identity of the LKP, is affected by boundary localized terms (BLTs) in the Lagrangian, which are not determined by the model itself. In minimal UED (MUED) models, it is assumed that all such terms vanish at the cutoff scale $\Lambda$ of the model, and are only generated at loop-level by renormalization group running. In the five-dimensional MUED model, the LKP is the first-level KK excitation of the U(1) gauge boson, the $B^1$ \cite{Cheng:2002iz}. The phenomenology of this DM candidate has been extensively investigated in the literature \cite{Servant:2002aq,Bergstrom:2004cy,Bergstrom:2004nr,Burnell:2005hm,Kong:2005hn,Arrenberg:2008wy,Belanger:2008gy,Belanger:2010yx,Cheng:2002ej,Bertone:2010fn}. See also the review in Ref.~\cite{Hooper:2007qk}.

In models beyond the MUED model, where the assumption that the BLTs vanish at the cutoff scale is relaxed, the particle masses and interactions generally change. In Ref.~\cite{Flacke:2008ne}, it was shown, using a restricted set of BLTs, that the first-level $Z$ boson, $Z^1$, or the first-level neutral Higgs boson, $H^1$, could be the LKP in such models.

The phenomenology of $Z^1$ DM has been studied in the literature. The relic density and direct detection prospects were considered in Ref.~\cite{Arrenberg:2008wy}. Neutrinos from $Z^1$ annihilations in the Sun were studied in Refs.~\cite{Flacke:2009eu,Blennow:2009ag} and the gamma ray spectrum from $Z^1$ annihilations in Refs.~\cite{Bonnevier:2011km,Melbeus:2011gs}. In general, the detection of $Z^1$ DM is more challenging than for the $B^1$, due to a larger preferred mass and a different distribution of annihilation channels. One exception might be the gamma ray line signal, which receives large contributions from the gauge boson self-interactions.

In this Letter, we study the remaining KKDM WIMP candidate, the $H^1$. In addition, the pseudo-scalar $A^{0,1}$ has exactly the same properties as $H^1$ from the DM point of view. Hence, our results hold also for that DM candidate. The rest of the Letter is organized as follows: In Sec.~\ref{sec:abun}, we calculate the relic abundance of $H^1$ DM, discussing also the effects of coannihilations. Then, in Sec.~\ref{sec:direct}, we estimate the direct detection prospects for this DM candidate. Next, in Sec.~\ref{sec:indirect}, we consider indirect detection through the photon line signal and continuum spectrum, positrons, and neutrinos from DM annihilations in the Sun. Finally, in Sec.~\ref{sec:conc}, we summarize our results and state our conclusions.

\section{\label{sec:abun}The relic abundance}

The standard calculation of the relic abundance of a thermally produced WIMP is reviewed, for example, in Ref.~\cite{Bertone:2010ww}. In the case that the mass splitting between the LKP and some of the other first-level KK modes is small, the effects of coannihilations are important \cite{Griest:1990kh}. Taking these effects into account, the abundance is given by
\begin{equation}\label{eq:Omega}
	\Omega h^2 \simeq \frac{1.04 \cdot 10^9~{\rm GeV}^{-1}}{M_{\rm Pl}} \frac{x_{\rm F}}{\sqrt{g_* (x_{\rm F})}} \frac{1}{I_a + 3 I_b / x_{\rm F}},
\end{equation}
where $x_{\rm F} = m_{\rm DM}/T_{\rm F}$, $m_{\rm DM}$ is the mass of the DM particle, $T_{\rm F}$ is the freeze-out temperature, $M_\text{Pl} \simeq 1.2 \cdot 10^{19}~{\rm GeV}$ is the Planck scale, and
\begin{equation}
	g_*(x_{\rm F}) = \sum_{i \in \left\{\text{bosons}\right\}} g_i + \frac{7}{8} \sum_{i \in \left\{\text{fermions}\right\}} g_i
\end{equation}
is the effective number of relativistic degrees of freedom at freeze-out. The quantities $I_a$ and $I_b$ are given by
\begin{eqnarray*}
	I_a \, & = & \, x_{\rm F} \int_{x_{\rm F}}^\infty a_{\rm eff} (x) x^{-2} \ud x,\\
	I_b \, & = & \, 2 x_{\rm F}^2 \int_{x_{\rm F}}^\infty b_{\rm eff} (x) x^{-3} \ud x.\\
\end{eqnarray*}
Here, $a_{\rm eff}$ and $b_{\rm eff}$ are defined by the expansion $\sigma_{\rm eff} = a_{\rm eff} + b_{\rm eff} v^2 + \mathcal{O} (v^4)$ of the effective cross section,
\begin{equation}
	\sigma_\text{eff} = \sum_{i,j} \sigma_{ij} \frac{g_i g_j}{g_\text{eff}^2} (1+\Delta_i)^{3/2} (1+\Delta_j)^{3/2} {\rm e}^{-x(\Delta_i + \Delta_j)},
\end{equation}
where $\sigma_{ij}$ is the coannihilation cross section between the states $i$ and $j$, $g_i$ is the number of degrees of freedom for the state $i$, and $\Delta_i = (m_i - m_\text{LKP}) / m_\text{LKP}$, with $m_{\rm LKP}$ denoting the mass of the LKP. Finally, the freeze-out temperature is obtained from the relation
\begin{equation}\label{eq:xF}
	x_{\rm F} = \ln \left[ c(c+2) \sqrt{\frac{45}{8}} \frac{g m_{\rm DM} M_{\rm Pl} (a_{\rm eff} + 6 b_{\rm eff} / x_{\rm F})}{2 \pi^3 \sqrt{g^* x_{\rm F}}} \right],
\end{equation}
where $g$ is the number of degrees of freedom for the DM particle, $c \simeq 1/2$ is determined numerically, and
\begin{equation}
	g_\text{eff}  = \sum_i g_i (1+\Delta_i)^{3/2} {\rm e}^{-x \Delta_i} \label{eq:geff}.
\end{equation}
In this work, we have used the software package micrOMEGAs \cite{Belanger:2010gh} to numerically calculate the relic density of $H^1$ DM, including coannihilations. In addition, we have checked the results analytically, using Eqs.~\eqref{eq:Omega}--\eqref{eq:geff}, and found agreement between the two methods.

\begin{figure}[t!]
\begin{center}
\includegraphics[width=.5\textwidth]{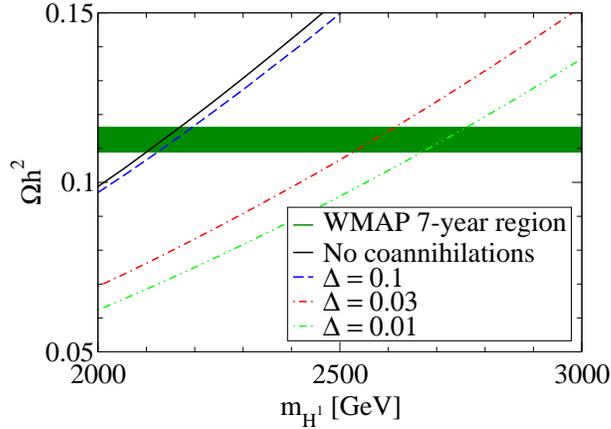}
\caption{The relic density of $H^1$ DM as a function of the mass $m_{H^1}$. The results are shown for the case without coannihilations as well as for coannihilations with a universal relative mass splitting $\Delta$ for all the first-level KK modes.} \label{fig:relicdensity}
\end{center}
\end{figure}

In the non-minimal UED model that we are studying, the mass spectrum has to be modified relative to the MUED model, so that $H^1$ is the LKP. Therefore, the rest of the first-level mass spectrum has to be fixed in some way in order to assess the effects of coannihilations, which depend strongly on the particle masses. Rather than considering the detailed effects of the BLTs, we take a more phenomenological approach by making an ansatz for the mass spectrum. Also, we do not consider modifications of the coupling constants due to the BLTs. The simplest ansatz for the mass spectrum is a universal mass splitting between the LKP and all the other first-level KK particles, parametrized by the relative mass splitting $\Delta = (m_1 - m_{\rm LKP})/m_{\rm LKP}$. This parametrization has previously been employed in the literature, \eg, in Refs.~\cite{Burnell:2005hm,Kong:2005hn}. The resulting relic density, as well as the relic density calculated without coannihilations, is presented in Fig.~\ref{fig:relicdensity}. We also show the 68~\% confidence region $\Omega h^2 = 0.1126 \pm 0.0036$ obtained from a combination of the WMAP seven-year data, baryon acoustic oscillations, and supernovae observations \cite{Komatsu:2010fb}. If coannihilations are not important, the $H^1$ relic abundance falls within this region for $m_{H^1} \simeq 2100~{\rm GeV}$. Coannihilations tend to increase this value, giving $m_{H^1} \simeq 2100~{\rm GeV}$ for $\Delta = 10~\%$, $m_{H^1} \simeq 2600~{\rm GeV}$ for $\Delta = 3~\%$, and $m_{H^1} \simeq 2700~{\rm GeV}$ for $\Delta = 1~\%$.

\begin{figure}[t!]
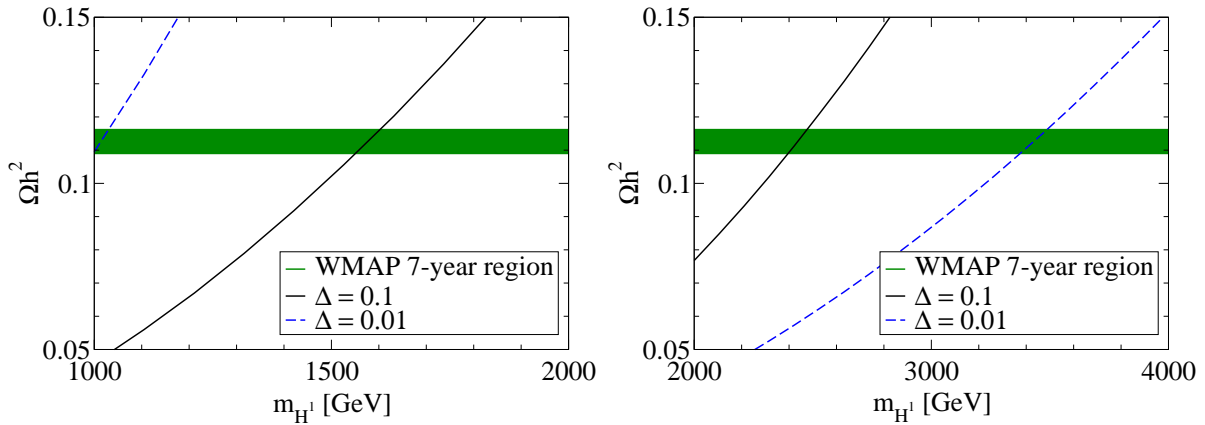

\begin{center}
\includegraphics[width=.49\textwidth]{relicdensity_leptons.eps}
\includegraphics[width=.49\textwidth]{relicdensity_quarks.eps}
\caption{The relic density of $H^1$ DM as a function of the mass $m_{H^1}$. Left panel: coannihilations with KK leptons only. Right panel: coannihilations with KK quarks only. The relative mass splitting for the coannihilating particles is given by $\Delta$.} \label{fig:relicdensity-separate}
\end{center}
\end{figure}

In Fig.~\ref{fig:relicdensity-separate}, we show the separate effects of coannihilations with leptons and quarks, respectively. We find that coannihilations with leptons tend to decrease the preferred mass, while coannihilations with quarks tend to increase it. In the most extreme case that we consider, \ie, $\Delta = 1~\%$, lepton coannihilations might shift the mass to about 1~TeV, while quark coannihilations could push it above 3~TeV.

\section{\label{sec:direct}Direct detection}

Now, we turn to the experimental signatures of $H^1$ DM, starting with direct detection experiments. The standard calculation procedure for WIMPs scattering on nuclei is reviewed in Ref.~\cite{Cerdeno:2010jj}. In addition to the WIMP-quark scattering cross sections, the scattering rate depends on the WIMP distribution in the vicinity of Earth as well as on the structure of the nucleons in terms of quarks and gluons. Throughout this Letter, we assume a Navarro--Frenk--White (NFW) halo profile \cite{Navarro:1995iw} with scale radius $r_{\rm S} = 20~{\rm kpc}$.

\begin{figure}[t!]
\begin{center}
\includegraphics[width=.4\textwidth]{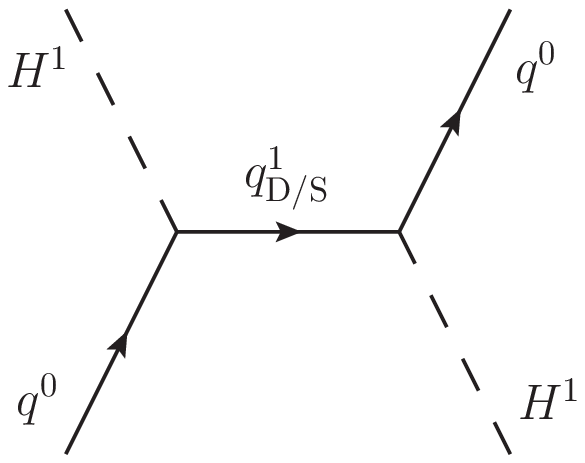}
\includegraphics[width=.4\textwidth]{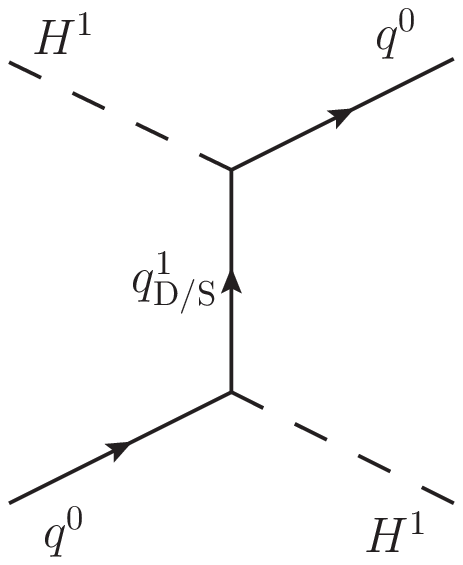}
\caption{Tree-level Feynman diagrams contributing to $H^1$ scattering on nuclei at quark level. The subscripts D and S on the KK quarks denote the SU(2) doublet and singlet Dirac fermions, respectively.} \label{fig:dirdet}
\end{center}
\end{figure}

The $H^1$, being a scalar particle, scatters only spin-independently on nuclei. The tree-level Feynman diagrams contributing to $H^1$-quark scattering are given in Fig.~\ref{fig:dirdet}.
The amplitude for the process $H^1 (p_1) q(p_2) \to H^1(p_3) q(p_4)$ is
\begin{equation}
	\mathcal{M} = -{\rm i} \frac{y_q^2}{2} \bar u_4 \left[ \frac{\slc{p}_1 + \slc{p}_2}{(p_1 + p_2)^2 - m_{q^1}^2} + \frac{\slc{p}_2 - \slc{p}_4}{(p_2 - p_4)^2 - m_{q^1}^2} \right] u_2,
\end{equation}
where $y_q$ is the Yukawa coupling of the quark flavor $q$ to the Higgs field, $u_i = u(p_i)$, and we have assumed $m_{q_{\rm D}^1} = m_{q_{\rm S}^1} \equiv m_{q^1}$. In the non-relativistic limit, $p_1 \simeq p_3 \simeq (m_{H^1}, \mathbf{0})$ and $\bar u_4 \gamma^0 u_2 = 2 m_q \xi_4^\dagger \xi_2$, while $\bar u_4 \gamma^i u_2 = 0$ for $i=1,2,3$. Expanding the amplitude to lowest order in $m_q$, we obtain
\begin{equation}
	\mathcal{M} = \mathcal{C}_q \xi_4^\dagger \xi_2,
\end{equation}
where
\begin{equation}
	\mathcal{C}_q = y_q^2 m_q \left[ \frac{1}{m_{H^1}^2 - m_{q^1}^2} + \frac{2 m_{H^1}^2}{(m_{H^1}^2 - m_{q^1}^2)^2} \right].
\end{equation}

The WIMP-nucleus cross section is given by
\begin{equation}
	\sigma_\text{SI} = \frac{1}{4\pi} \frac{\mu^2}{m_{H^1}^2} \left[ Z f^p + (A-Z) f^n \right]^2,
\end{equation}
where the reduced mass $\mu = m_{H^1} m_N / (m_{H^1} + m_N)$, $m_N$ is the mass, $Z$ the atomic number, and $A$ the mass number of the nucleus, and
\begin{equation}\label{eq:fpn}
	f^{p,n} = \sum_q \mathcal{C}_q \langle p | \bar q q | p \rangle = m_p \left[ \sum_{q=u,d,s} \frac{\mathcal{C}_q}{m_q} f_{T_q}^{p,n} + \frac{2}{27} f_{TG}^{p,n} \sum_{q=c,b,t} \frac{\mathcal{C}_q}{m_q} \right].
\end{equation}
Here, $f_{T_u}^p = 0.020 \pm 0.004$, $f_{T_u}^n = 0.014 \pm 0.003$, $f_{T_d}^p = 0.026 \pm 0.005$, $f_{T_d}^n = 0.036 \pm 0.008$, $f_{T_s}^{p,n} = 0.118 \pm 0.062$, $f_{TG}^p = 1 - f_{T_u}^p - f_{T_d}^p - f_{T_s}^p \simeq 0.84$, and similarly $f_{TG}^n \simeq 0.83$ \cite{Ellis:2000ds}.

Since the amplitude for the contribution from the quark flavor $q$ is proportional to the square of the Yukawa coupling, $y_q^2 \propto (m_q / v)^2$, where $v$ is the vacuum expectation value of the Higgs field, the scattering is suppressed for all flavors except for the top quark. The heavy quarks $q = c,b,t$ contribute to the scattering only through loop-level couplings to gluons. The effective couplings for these quarks, given in Eq.~\eqref{eq:fpn}, were first derived in Ref.~\cite{Drees:1993bu} for neutralino-nucleon scattering, and they do not hold in general. Nevertheless, we use these expressions to estimate the contributions from the top quark, as the contributions from the light quarks are completely negligible for the $H^1$.

Experimental results are usually expressed in terms of the WIMP-nucleon cross section $\sigma_n = \sigma_\text{SI} m_p^2 / (\mu^2 A^2)$. Using $f_{TG}^{\rm p} \simeq f_{TG}^{\rm n}$, we obtain
\begin{equation}
	\sigma_n \simeq \frac{1}{2916 \pi} (f_{TG}^{\rm p})^2 y_t^2 \frac{1}{\Delta_q^4} \left( \frac{m_p}{m_{H^1}} \right)^4 \frac{1}{m_{H^1}^2} \simeq (6 \cdot 10^{-10}~{\rm pb}) \left( \frac{2~{\rm TeV}}{m_{H^1}} \right)^6 \left( \frac{0.03}{\Delta_q} \right)^4,
\end{equation}
where $\Delta_q = (m_{q^1} - m_{H^1})/m_{H^1}$. For $\Delta_q = 3~\%$ and $m_{H^1} = 2~{\rm TeV}$, the scattering cross section is several orders of magnitude below the sensitivities of current direct detection experiments, such as XENON100 \cite{Aprile:2011hi}, in the relevant mass range. The cross section increases with decreasing mass $m_{H^1}$ and/or mass splitting $\Delta_q$. However, coannihilations with KK quarks drive the mass to larger values, as observed in Fig.~\ref{fig:relicdensity-separate}. Hence, if $\Delta_q$ is small, there is a tendency to drive the mass $m_{H^1}$ to a value that is too large for successful direct detection.

\section{\label{sec:indirect}Indirect detection}

Next, we turn to the indirect detection of $H^1$ DM through the observations of decay products from $H^1$ pair annihilations. We consider high-energy photons, positrons, and neutrinos from $H^1$ annihilations in the Sun. The model-dependent input to the indirect detection signals is the total annihilation cross section and the branching ratios into different final states. Since the typical WIMP velocity is $v \simeq 10^{-3}$, these quantities are calculated at zero momentum, \ie, only $s$-wave contributions are taken into account. The $H^1$ annihilates dominantly into the final states $HH$, $ZZ$, and $W^+ W^-$, and the total annihilation cross section is given by  
\begin{equation}
	\sigma v (H^1 H^1) \simeq (0.83~{\rm pb}) \left( \frac{2~{\rm TeV}}{m_{H^1}} \right)^2.
\end{equation}
For scalar DM, annihilations into fermion-antifermion pairs are helicity suppressed, \ie, the $s$-wave cross sections are proportional to $(m_f/m_{f^1})^2$. Since the relic abundance requires a relatively large value for $R^{-1}$, the annihilation cross sections are small even for the top quark. The branching ratios into all available SM final states are given in Table~\ref{tbl:branchingratios}, computed in the limit of a degenerate first-level KK mass spectrum.
\begin{table}
\begin{center}
\begin{tabular}{|l|l|}
	\hline
	\textbf{Final state} & \textbf{Branching ratio} \\
	\hline
	$H H$ 		& 0.543 \\
	$Z Z$ 		& 0.237 \\
	$W^+ W^-$ 	& 0.220 \\
	$\bar f f$ 	& 0 \\
	\hline
\end{tabular}
\caption{Branching ratios into all final state channels for $H^1$ DM. The branching ratios are computed in the limit of degenerate first-level KK masses. Annihilation into fermion-antifermion pairs is helicity suppressed, and thus, it is negligible due to the large compactification scale preferred for the relic abundance.}\label{tbl:branchingratios}
\end{center}
\end{table}

\subsection{\label{sec:photon-peak}The gamma ray line signal}

A smoking-gun signature of DM would be the detection of a gamma ray line signal, coming from the loop-level process $H^1 H^1 \to \gamma \gamma$, with $E_\gamma = m_{H^1}$. In addition, the processes $H^1 H^1 \to \gamma Z, \gamma H$ could give line signals at $E_\gamma = m_{H^1} [ 1 - m_X^2 / (4 m_{H^1}^2) ]$, where $X = Z, H$. The relative shift from the two-photon peak, $\Delta E_\gamma = - m_X^2 / (4 m_{H^1}^2) \simeq 10^{-3}$, is too small to be resolved experimentally, and hence, the individual peaks add up to a single one. For $H^1$ DM, $s$-wave annihilation into the $\gamma H$ final state is not possible due to conservation of angular momentum, and therefore, the process is suppressed. The $H^1 H^1 \to \gamma Z$ process is discussed below.

The amplitude for the process $H^1 (p_1) H^1 (p_2) \to \gamma (p_3) \gamma (p_4)$ can be written as
\begin{equation}
 \mathcal{M} = \mathcal{M}_{\mu\nu} \epsilon_3^{*\mu} (p_3) \epsilon_4^{*\nu} (p_4),
\end{equation}
where $\epsilon_i^*$ are the photon polarization tensors. For annihilation at rest, $p_1 \simeq p_2 \simeq p = (m_{H^1},\mathbf{0})$. Using conservation of 4-momentum, $2p + p_3 + p_4 = 0$, the transversality of the polarization tensors, $p_i \cdot \epsilon_i^* = 0$, and the Ward identity $p_{3\mu} M^{\mu\nu} = p_{4\nu} M^{\mu\nu} = 0$, the tensor $\mathcal{M}^{\mu\nu}$ can be reduced to the simple form
\begin{equation}
	\mathcal{M}^{\mu\nu} = B \left( \frac{p_3^\nu p_4^\mu}{m_{H^1}^2} - 2 g^{\mu\nu} \right),
\end{equation}
where the quantity $B$ depends on the particle masses only. The cross section is given by
\begin{equation}
	\sigma v =  \frac{|B|^2}{8 \pi m_{H^1}^2}.
\end{equation}

\begin{figure}[t!]
\begin{center}
\includegraphics[width=.3\textwidth]{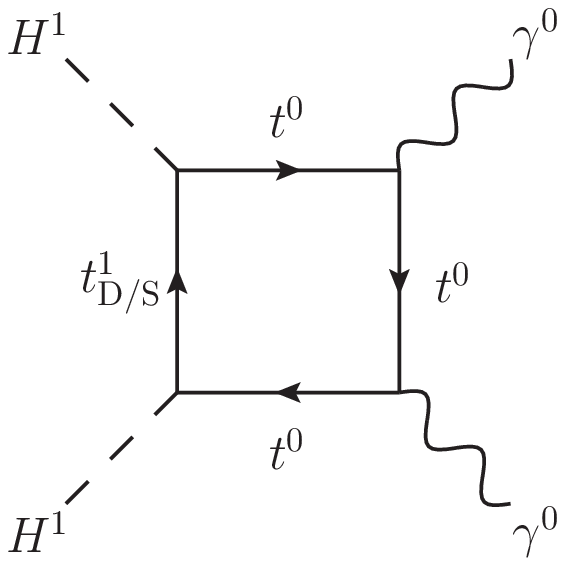}
\includegraphics[width=.3\textwidth]{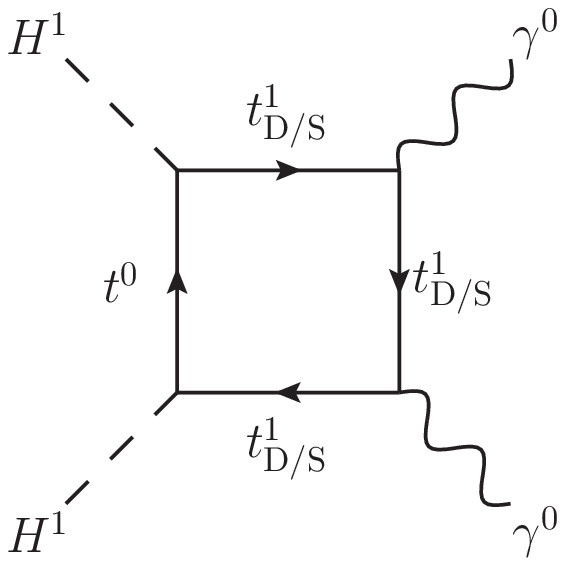}
\includegraphics[width=.3\textwidth]{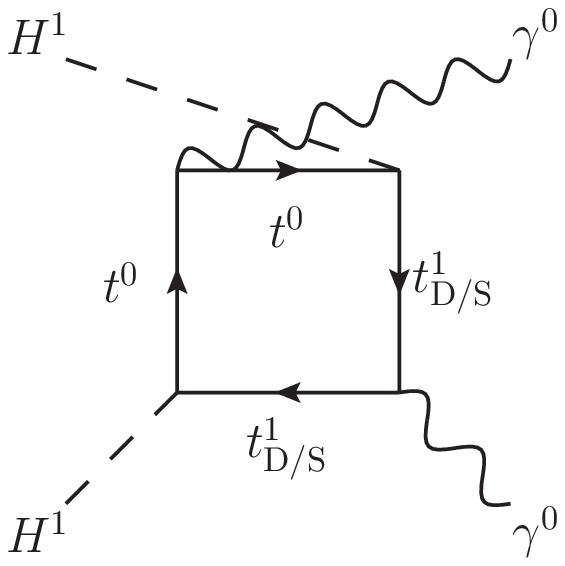}
\caption{One-loop top quark mediated Feynman diagrams contributing to the process $H^1 H^1 \to \gamma \gamma$. In addition to these diagrams, there are three diagrams with crossed final state particles. The subscripts D and S on the KK quarks denote the SU(2) doublet and singlet Dirac fermions, respectively.} \label{fig:monogamma}
\end{center}
\end{figure}

In Fig.~\ref{fig:monogamma}, we show the Feynman diagrams for $H^1 H^1 \to \gamma \gamma$ that involve internal top quarks. From an analysis of the coupling constants only, we would expect these to be larger than the corresponding diagrams that involve internal bosons by a factor $(y_t / g)^4 \simeq 5$.
From this subset of diagrams, we find that
\begin{equation}
	B = 8 \eta C_0 (0,0,4,\eta,\eta,\eta) - 4 C_0 (1,0,-1,0,\eta,\eta) - 4,
\end{equation}
where $\eta = (m_{t^1}/m_{H^1})^2$ and
\begin{eqnarray}
	C_0 (0,0,4,\eta,\eta,\eta) \, & = & \, -\frac{1}{2} \arctan^2 \left( \frac{1}{\sqrt{\eta-1}} \right), \\
	C_0 (1,0,-1,0,\eta,\eta) \, & = & \, \frac{1}{2} \left[ {\rm Li}_2 \left( -\frac{1}{\eta} \right) - {\rm Li}_2 \left( \frac{1}{\eta} \right) \right].
\end{eqnarray}
Here, ${\rm Li}_2 (x)$ denotes the dilogarithm, 
\begin{equation}
	{\rm Li}_2 (x) = - \int_0^1 \frac{\log(1-xt)}{t} \ud t.
\end{equation}

\begin{figure}[t!]
\begin{center}
\includegraphics[width=.5\textwidth]{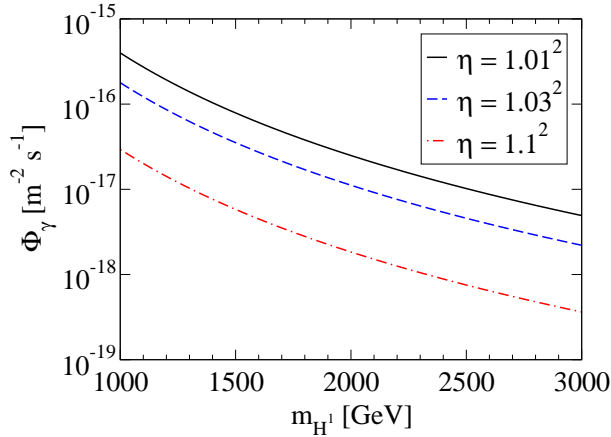}
\caption{The integrated gamma ray flux from the process $H^1 H^1 \to \gamma \gamma$ as a function of the LKP mass $m_{H^1}$, for different values of the mass splitting $\eta = (m_{t^1}/m_{H^1})^2$.} \label{fig:monogamma-flux}
\end{center}
\end{figure}

Finally, the flux at Earth in the direction of the galactic center is given by \cite{Bergstrom:1997fj}
\begin{equation}
	\Phi_\gamma \simeq (4.7 \cdot 10^{-12}~{\rm m}^{-2} \ {\rm s}^{-1}) \left( \frac{\sigma v }{10^{-29}~{\rm cm}^3 \ {\rm s}^{-1}} \right) \left( \frac{2~{\rm TeV}}{m_{H^1}} \right)^2 \langle J_\text{GC} \rangle_{\Delta \Omega} \Delta \Omega,
\end{equation}
where the solid angle $\Delta \Omega$ represents the resolution of the detector and $\langle J_\text{GC} \rangle_{\Delta \Omega}$ is the dimensionless line-of-sight integral in the direction of the galactic center. For $\Delta \Omega = 10^{-5}$, $\langle J_\text{GC} \rangle_{\Delta \Omega} \Delta \Omega \simeq 0.13$ for the NFW halo profile. In Fig.~\ref{fig:monogamma-flux}, we present the numerical results for a number of different values for the mass-splitting parameter $\eta$. In addition, we have calculated the contribution from diagrams involving internal bosons numerically, and found a result which is of the same order of magnitude as that from the top quark diagrams. Hence, the order of magnitude of our result is correct. Due to the small size of the line signal, however, a more detailed calculation would not be useful.

We have also studied the $\gamma Z$ final state, and found that the corresponding photon flux is smaller than from the $\gamma \gamma$ final state by about one order of magnitude. This is partly due to the difference in couplings, and partly due to the fact that each $H^1 H^1 \to \gamma Z$ annihilation only results in a single photon.

\subsection{\label{sec:photon-continuum}The photon continuum spectrum}

\begin{figure}[t!]
\begin{center}
\includegraphics[width=.5\textwidth]{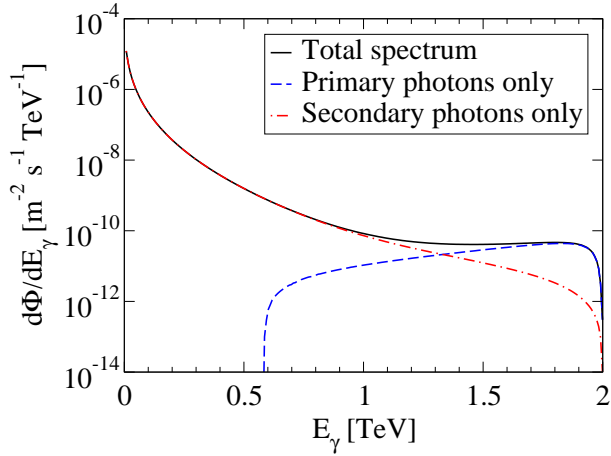}
\caption{The differential photon spectrum as a function of $E_\gamma$, for $m_{H^1} = 2~{\rm TeV}$.} \label{fig:gammadiff}
\end{center}
\end{figure}

In addition to the line signal, there is a continuous component of the photon spectrum. Primary photons are produced in final state radiation (FSR) processes, \ie, three-body processes of the form $H^1 H^1 \to X \bar X \gamma$, where $X$ denotes an electrically charged SM particle. These processes are suppressed by a factor $\alpha \simeq 1/137$ relative to two-body annihilation processes. However, the addition of a photon in the final state opens up the possibility of annihilations into fermion-antifermion pairs, which are otherwise helicity suppressed. The situation is similar to the case of neutralino annihilations, where the Majorana nature of the neutralino leads to helicity suppression. In addition to primary photons, secondary photons are produced in the decays of other final states, which are directly produced, \eg, quarks, leptons, and gauge bosons. Both of these contributions are calculated in micrOMEGAs. In Fig.~\ref{fig:gammadiff}, we present the total continuum spectrum as well as the separate contributions from primary and secondary photons. The hard part of the spectrum is dominated by the FSR. In contrast to the case of $B^1$ annihilations, there is no sharp cutoff at $E_\gamma = m_{H^1}$, due to the different distribution of annihilation products. The secondary photons generated by the decays of other particles mainly contribute to the soft end of the spectrum.

The differential flux close to the endpoint $E_\gamma = m_{H^1}$ is larger than the line signal by several orders of magnitude. Hence, the line signal is completely negligible in comparison to the continuum spectrum.

\subsection{\label{sec:PosProt}Positrons}

In general, positrons from DM annihilations can be produced directly through the $e^+ e^-$ annihilation channel as well as indirectly through the decays of other annihilation products. For the $H^1$, the direct $e^+ e^-$ channel is suppressed by the small Yukawa coupling constants for light fermions. Hence, the spectrum is dominated by secondary positrons, and therefore, it is relatively soft.

The observed flux at Earth depends strongly on the propagation of the positrons through the galaxy. The most important effects are space diffusion and energy losses due to synchrotron radiation and inverse Compton scattering \cite{Salati:2010}. In micrOMEGAs, the propagation is modeled as a diffusion-loss equation for the number density of positrons per unit volume and energy, $\psi_{e^+} = \ud n_{e^+} / \ud E$, which is of the form
\begin{equation}
	- \nabla \cdot \left[ K(E) \nabla \psi_{e^+} \right] - \frac{\partial}{\partial E} \left[ b(E) \psi_{e^+} \right] = Q_{e^+} (\mathbf{x},E).
\end{equation}
Here, $Q_{e^+}$ is the source term, the space diffusion coefficient $K = K_0 (E / E_0)^{0.7}$ with $K_0 = 0.0112~{\rm kpc}^2 / {\rm Myr}$ and $E_0 = 1~{\rm GeV}$, and the positron loss rate $b(E) = E^2 / (E_0 \tau_E)$ with the energy loss time $\tau_E = 10^{16}~{\rm s}$.

\begin{figure}[t!]
\begin{center}
\includegraphics[width=.5\textwidth]{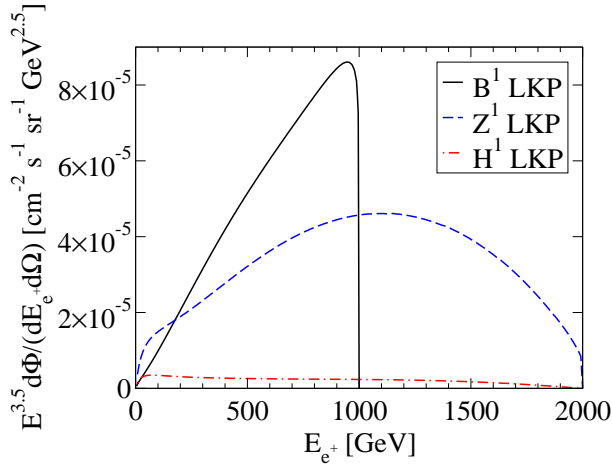}
\caption{The differential positron spectra $E^{3.5} \ud \Phi / (\ud E_{e^+} \ud \Omega)$ for $B^1$, $Z^1$, and $H^1$ DM annihilations, with $m_{B^1} = 2~{\rm TeV}$, $m_{Z^1} = 2~{\rm TeV}$, and $m_{B^1} = 1~{\rm TeV}$, which are typical values for the respective LKPs if the correct DM abundance should be obtained.} \label{fig:posidiff}
\end{center}
\end{figure}

The positron flux at Earth is shown in Fig.~\ref{fig:posidiff} for $m_{H^1} = 2~{\rm TeV}$. For comparison, we also give the corresponding results for annihilations of $B^1$ and $Z^1$ DM, for which we have chosen the typical masses $m_{B^1} = 1~{\rm TeV}$ and $m_{Z^1} = 2~{\rm TeV}$ given by the relic abundance calculations for the respective LKP candidates. The $B^1$ spectrum has been shown to be possible to fit to the PAMELA data, although the predicted magnitude is too small, requiring a boost factor of about $10^3$ \cite{Hooper:2009fj}. The $H^1$ spectrum is relatively soft and has no sharp cutoff at $E_{e^+} = m_{H^1}$. In comparison to the flux from $B^1$ annihilations, it is smaller by about two orders of magnitude.

\subsection{\label{sec:neutrino}The neutrino spectrum}

The gravitational capture of WIMPs in the Sun could lead to a significantly enhanced density, giving neutrino signals from pair annihilations. However, the capture rate of WIMPs in the Sun is proportional to the WIMP-proton cross section, which is the same quantity that is constrained by direct detection experiments. This means that the expected results in neutrino telescopes are correlated with the limits from direct detection experiments. An important consequence is that for DM candidates that interact only spin-independently with nuclei, the constraints from direct detection experiments are already strong enough to rule out an observation of neutrinos from DM annihilations in the Sun \cite{Halzen:2005ar}. This is the case for scalar DM candidates, such as the $H^1$, and hence, neutrinos from the Sun are not a promising detection channel for this DM candidate.

\section{\label{sec:conc}Summary and conclusions}

In this Letter, we have investigated the first-level KK excitation of the Higgs boson in non-minimal UED models as a DM candidate. We have calculated the relic abundance, including coannihilations with other first-level KK particles. In addition, we have studied the detection prospects in direct as well as indirect DM detection experiments.

Although the $H^1$ is a typical WIMP DM candidate, we find that detecting it would be very challenging in direct as well as indirect detection experiments. The main reasons for the suppressed rates are the facts that the Yukawa couplings are small for all fermions except for the top quark and that annihilation into any fermion-antifermion pair is helicity suppressed. This means that the $H^1$-quark coupling relevant for direct detection is small, that the continuum gamma ray spectrum does not display a sharp cutoff, and that the positron spectrum is soft, consisting mainly of secondary positrons. In addition, being a scalar, the $H^1$ interacts only spin-independently with nuclei, and is not expected to give observable signatures in neutrinos from the Sun.

To conclude, the DM phenomenology of the $H^1$ is very different from that of the standard KKDM candidate, the $B^1$. The $B^1$ has relatively good detection prospects, especially in indirect detection experiments. The photon and positron channels both feature hard spectra with sharp cutoffs at $E_{\gamma,e^+} = m_{H^1}$, and also, the mainly spin-dependent $B^1$-nucleon interactions give rise to potentially strong signatures in neutrinos from the Sun. The $H^1$ phenomenology is actually more similar to that of the $Z^1$, which has a similar preferred mass range for the relic abundance, and also features large branching ratios into bosons, rather than fermions. The main exception is the gamma ray line signal, which is expected to be strong for the $Z^1$, due to large contributions from the non-Abelian gauge boson self-interactions, but not for the $H^1$.

It is apparently extremely hard to positively identify the $H^1$ as a DM particle. However, it would in principle be possible to produce it at a collider like the LHC. If collider experiments revealed part of the KK spectrum, with an $H^1$-like particle as the LKP, while direct and indirect DM detection experiments only kept setting limits, then one would be very tempted to declare the $H^1$ as the DM particle. A more detailed experimental investigation of its properties could then be used to probe the agreement of abundance calculations with the observed value. In short, although the experimental situation is very challenging, there is still hope to be able to establish the $H^1$ as DM in the future.

Throughout the Letter, we have assumed that the coupling constants are not affected by the BLTs. In general, this might not be the case, and the results could be changed by such effects. An investigation of this issue is beyond the scope of this work.

Finally, we repeat that the results of this Letter can be directly carried over to the pseudo-scalar $A^{0,1}$.

\section*{\label{sec:ack}Acknowledgments}

This work was supported by the Swedish Research Council (Vetenskapsr{\aa}det), contract no.\ 621-2011-3985 (T.O.), and by the G\"oran Gustafsson Foundation (A.M.). A.M.~is now supported by a Marie Curie Intra-European Fellowship within the 7th European Community Framework Programme FP7-PEOPLE-2011-IEF, contract PIEF-GA-2011-297557.



\end{document}